\documentclass[9pt,conference]{IEEEtran}

\usepackage{waspaa25}

\usepackage{bm} 
\usepackage{hyperref}
\usepackage{enumitem}
\usepackage{booktabs}
\usepackage{siunitx}
\usepackage{threeparttable}
\usepackage{makecell}
\usepackage{pifont}
\usepackage{enumitem}
\usepackage{xcolor}
\usepackage{comment}
\usepackage{amsmath}  
\usepackage{array}    
\usepackage{multirow,multicol}
\usepackage{float}

\newcommand{\yes}{\ding{51}} 
\newcommand{\no}{\ding{55}} 
\newcommand{\maybe}{$\sim$}     
\title{Benchmarking Sub-Genre Classification For Mainstage Dance Music}

\name{Hongzhi Shu\textsuperscript{1*}, Xinglin Li\textsuperscript{2*}, Hongyu Jiang\textsuperscript{3}, Minghao Fu\textsuperscript{4}, Xinyu Li\textsuperscript{4†}}
\address{\textsuperscript{1}Johns Hopkins University, Baltimore, USA \\
        \textsuperscript{2}Southeast University, Nanjing, China \\
        \textsuperscript{3}National University of Defense Technology, Changsha, China \\
        \textsuperscript{4}Mohamed bin Zayed University of Artificial Intelligence, Abu Dhabi, UAE \\
        \thanks{\textsuperscript{*}Equal contribution.}
\thanks{\textsuperscript{†}Corresponding author: xinyu.li@mbzuai.ac.ae.}}

\makeatletter
\renewcommand{\subsubsection}{\@startsection{subsubsection}{3}{\z@}%
  {-3.25ex\@plus -1ex \@minus -.2ex}
  {1.5ex \@plus .2ex}
  {\normalfont\small\bfseries}}
\makeatother

\begin{document}
\maketitle
\begin{abstract}
Music classification, a cornerstone of music information retrieval, supports a wide array of applications. To address the lack of comprehensive datasets and effective methods for sub-genre classification in mainstage dance music, we introduce a novel benchmark featuring a new dataset and baseline. Our dataset expands the scope of sub-genres to reflect the diversity of recent mainstage live sets performed by leading DJs at global music festivals, capturing the vibrant and rapidly evolving electronic dance music (EDM) scene that engages millions of fans worldwide. We employ a continuous soft labeling approach to accommodate tracks blending multiple sub-genres, preserving their inherent complexity. Experiments demonstrate that even state-of-the-art multimodal large language models (MLLMs) struggle with this task, while our specialized baseline models achieve high accuracy. This benchmark supports applications such as music recommendation, DJ set curation, and interactive multimedia systems, with video demos provided. Our code and data are all open-sourced at \textit{\href{https://github.com/Gariscat/housex-v2.git}{https://github.com/Gariscat/housex-v2.git}}.

\end{abstract}
%
%
\section{Introduction}
\label{sec:intro}
Being one of the most fundamental and important tasks in music information retrieval (MIR), music genre classification has mainly focused on broad genres \cite{1021072, fma_dataset, fma_challenge, Zhou_2025}. Despite continuous progress, existing datasets lack fine-grained labels that can capture the nuances within electronic dance music (EDM), and with 0/1 labels, they hardly manifest the category overlap in EDM data. Moreover, current universal models have subpar performance in specific tasks \cite{Qwen-Audio,chu2024qwen2audiotechnicalreport,xu2025qwen25omnitechnicalreport,kimiteam2025kimiaudiotechnicalreport}. The limitations are evident in the contexts of mainstage DJ sets, where tracks usually fall into sub-genres within the broad category of house music. Hence, the unique challenges of EDM necessitate specialized datasets and algorithms that are tailored to its structural characteristics and the complexity of its production techniques.

To address this gap, we introduce a new benchmark specifically targeting at the classification of house music sub-genres. A dataset is designed to provide annotations from a list of 8 sub-genres. These 8 classes were selected to represent a comprehensive range of prominent sub-genres, reflecting their popularity and distinctive characteristics that highlight the diversity and evolution of EDM shaped by the influence of music festivals and digital media. Unlike existing dataset \textit{HouseX} \cite{9980316}, we introduce soft labeling instead of 0/1 categorical labeling to provide detailed and nuanced representation of the music. In addition, a baseline model is presented using spectral features \cite{Schrkhuber2010CONSTANTQTT, Schrkhuber2014AMT}, building a foundation for future research in dance music genre classification. Lastly, we develop a prototype of automated music visualization based on our proposed model, demonstrating its potential to enhance the visual experience.

Overall, this work aims to advance MIR for mainstage dance music by offering data, models, and a demo application. Our key contributions include: (a) improved annotation paradigm with an extended dataset; (b) strong specialized models and experiment results where MLLMs fail; (c) a multimedia application demo (visual FX automation).


\begin{table*}[!htbp]
\centering
\scriptsize
\caption{Determinants of sub-genres, where $x$-$y$ indicates the lowerbound $x$ and the upperbound $y$ of the attribute, voted by human musicians/producers.}
\label{tab:house_music_genres}
\begin{threeparttable}
    \begin{tabular}{l||p{1.8cm}p{1.8cm}p{1.8cm}p{1.8cm}p{1.8cm}p{1.8cm}p{1.8cm}}
    \hline
    \textbf{Genre} & \textbf{Lead Inst.} & \textbf{Chord Inst.} & \textbf{Bass Inst.} & \textbf{Groove} & \textbf{Rhythm} & \textbf{Distortion} & \textbf{Organicity} \\
    \hline
    Progressive House & \yes & \yes & \yes & 4-5 & 1-3 & 1-2 & 3-5 \\
    Future House      & \yes & \maybe & \yes & 3-4 & 4-5 & 3-4 & 1-3 \\
    Bass House        & \yes & \no    & \yes & 2-4 & 3-4 & 4-5 & 1-2 \\
    Tech House        & \no  & \no    & \yes & 3-5 & 3-4 & 2-4 & 1-3 \\
    Deep House        & \yes & \maybe & \yes & 2-3 & 1-2 & 1-2 & 2-4 \\
    Bigroom           & \yes & \no    & \no  & 1-2 & 1-3 & 4-5 & 1-2 \\
    Future Rave       & \yes & \no    & \yes & 4-5 & 3-5 & 2-4 & 1-2 \\
    Slap House        & \yes & \no    & \yes & 3-4 & 3-5 & 1-3 & 1-2 \\
    \hline
    \end{tabular}
    \begin{tablenotes}
        \item [a] \yes\ denotes ``yes", \no\ denotes ``no", and $\sim$ denotes ``uncertain", which could be either yes or no depending on the track.
    \end{tablenotes}
\end{threeparttable}
\end{table*}

\begin{figure*}[!htp]
    \begin{center}
        \includegraphics[width=0.92\textwidth]{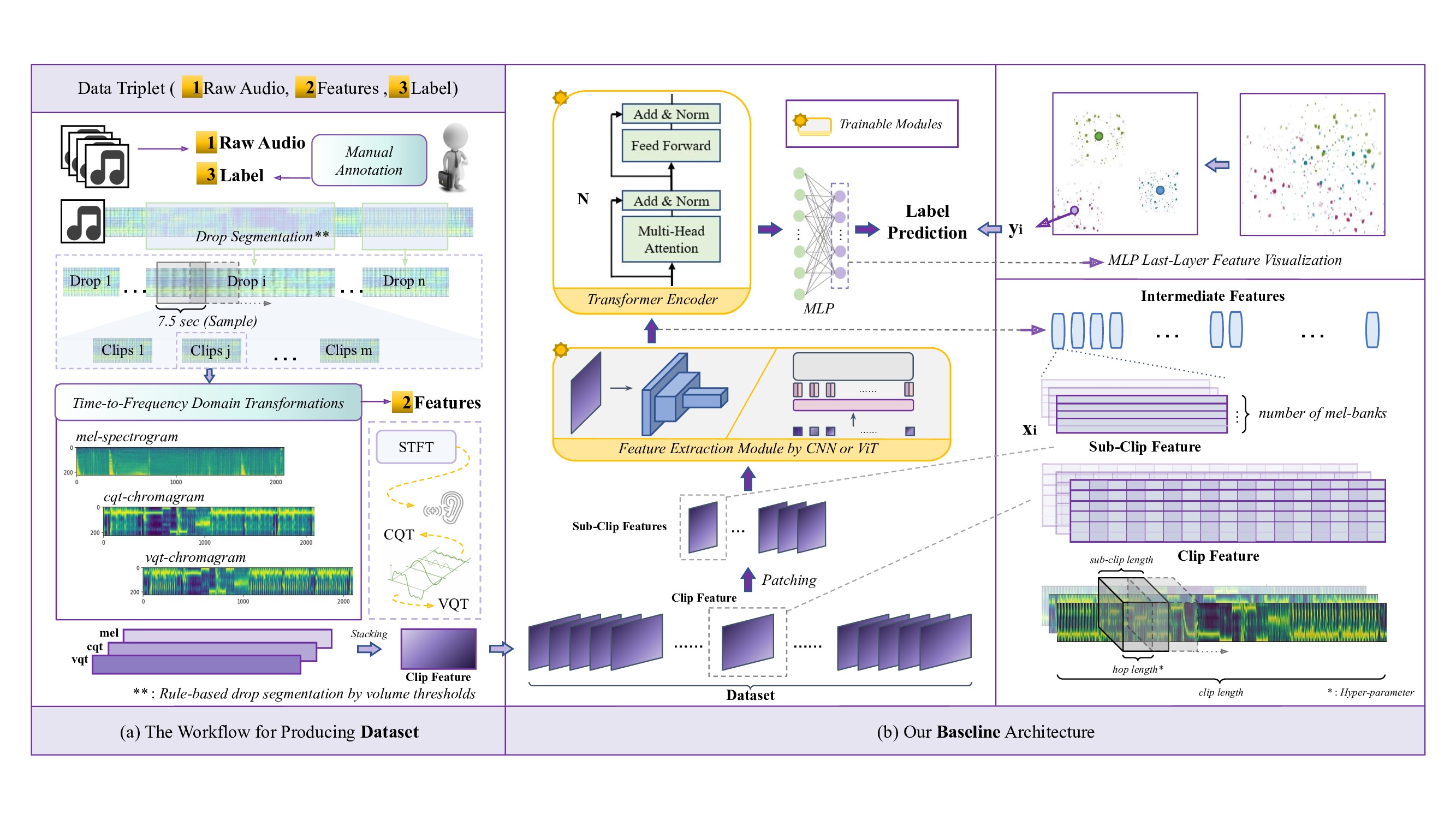} 
        \caption{Pipeline Structure from Data Collection to Model Design}
        \label{fig:pipeline}
    \end{center}
\end{figure*}

\section{Related Work}
\label{sec:related-work}

Music information retrieval (MIR) has been a research focus since the early 21st century, as evidenced by several comprehensive surveys \cite{Survey2005, 8187204, Simonetta_2019}. Traditional music genre datasets, like \textit{GTZAN} \cite{1021072}, \textit{FMA} \cite{fma_dataset} and \textit{MSD} \cite{bertin2011million}, often focus on broad genres like pop, country, and rock. Other general datasets like \textit{MusicCaps} \cite{agostinelli2023musiclmgeneratingmusictext} provide music-text pairs from which one can extract genre information. The \textit{HouseX} dataset \cite{9980316} advanced classification in EDM but faced challenges like limited category richness and scale. Prior works \cite{Caparrini26052020,hsu2021deeplearningbasededm} studied EDM sub-genre classification by traditional methods and deep learning, respectively. In addition, previous works tackling drop\footnote{In electronic dance music, the ``drop'' refers to a structural climax, typically marked by an abrupt increase in sonic intensity, rhythmic complexity, and spectral energy, serving a role comparable to the chorus in popular music.} detection \cite{argüello2024cuepointestimationusing,DBLP:conf/ismir/YadatiLLH14,Huang_2018} also highlight the emerging research interest in MIR for EDM. Alongside traditional deep learning methods, recent progress in language models for music understanding like \cite{liu2023music} and \cite{deng2024musilingo}, and strong multimodal large language models (MLLMs) like \cite{Qwen-Audio,chu2024qwen2,xu2025qwen25omnitechnicalreport,kimiteam2025kimiaudiotechnicalreport} could also be used to classify music audio.

This work presents a new dataset expanding from a smaller dataset \cite{9980316}, featuring 1000+ production-level tracks from top record labels that span across 8 sub-genres. To the best of our knowledge, given that \cite{hsu2021deeplearningbasededm} only open-sourced their model checkpoints, our work is the only publicly available large-scale dataset for EDM sub-genre classification. We use soft labeling to capture tracks with multiple sub-genre traits\footnote{Please refer to the code repository for an example.}. Our architecture employs convolutional neural networks (CNNs) \cite{He2015DeepRL,Huang2016DenselyCC,Simonyan2014VeryDC} and vision transformers (ViTs) \cite{Dosovitskiy2020AnII} to extract features with a sliding window, followed by some transformer encoder layers to predict the target distribution. Such customized models for sub-genre classification significantly outperform the existing MLLMs that take audio as direct input.

\section{Methodology}
\label{sec:methodology}

This section outlines the methodology used in our study, encompassing the definition and annotation of the dataset, as well as the training of the classification model. The complete pipeline, from data preparation to model training/inference, is illustrated in Figure~\ref{fig:pipeline}. For simplicity and clarity, we refer to sub-genres of house music as ``genres" throughout this section unless otherwise specified.

\subsection{Dataset}
\label{sub-sec:benchmark}
EDM encompasses a diverse range of sub-genres, each characterized by distinct rhythmic, timbral, and structural features. While these sub-genres are well-known within the EDM community, they may be unfamiliar to broader audiences in audio signal processing and music information retrieval. To facilitate understanding of our dataset and classification methodology, we first define the eight EDM sub-genres considered in this study, highlighting their key characteristics and relevance to our analysis, followed by the collection and labeling process.

\subsubsection{Definitions}
We provide both qualitative and quantitative descriptions of the 8 genres. A key observation is that the values of soft labels are \underline{\textbf{not}} absolute per se. Even for professionals, one track can be interpreted in multiple ways. For example, track $X$ is generally classified as genre $A$ but also exhibits characteristics of genre $B$. Therefore, a soft label can better preserve this nuance than a binary 0/1 label that assigns the track solely to genre $A$. In Table~\ref{tab:house_music_genres}, we show how different sub-genres are characterized by the following aspects.

\begin{enumerate}[label=\roman*]
    \item The first three columns (\textbf{xxx. Inst.}) indicate whether the track specifically assigns instruments for the parts, respectively. For example, some future house tracks do not use chord instruments since their lead sounds can cover the mid frequencies.
    \item \textbf{Groove} measures the energy fluctuation of the instrumental part (no percussion) within a beat, from 1 (very present, extremely short attack) to 5 (highly-sidechained).
    \item \textbf{Rhythm} discusses how fragmented/off-beat the notes are. We consider the length of notes played by the instruments in the track. 1 means extremely coarse (mainly 1/4 notes and 1/8 notes) and 5 means extremely fine (1/32 notes exist).
    \item \textbf{Distortion} describes the extent to which the track uses ``dirty" and ``inharmonic" sounds/effects, such as clipping, flanging, etc. 1 is very clean (hardly any noticeable distortion) and 5 is very harsh (almost distortion in every bar). 
    \item \textbf{Organicity} measures the extent to which the track uses organic instruments like acoustic piano/guitar/strings, etc.
\end{enumerate}

We also provide plain-English descriptions, including genre-specific characteristics (e.g., vocal leads, plucked bass) and one representative example for each sub-genre. Due to space constraints, these are included in the code repository.

\subsubsection{Annotation}

The input data are represented as triplets, each comprising the raw audio (\(X_{\text{raw}}\)), its extracted acoustic features (\(X_{\text{feat}}\)), and a corresponding label (\(Y\)), as illustrated by numbers from 1 to 3 in Figure~\ref{fig:pipeline}. Note that our model does not take in audio waveforms, which are used as input by MLLMs such as \textit{Qwen-Audio} \cite{Qwen-Audio} and \textit{Kimi-Audio} \cite{kimiteam2025kimiaudiotechnicalreport} that include built-in spectrogram extractors or other audio encoders like \textit{Whisper} \cite{openai-whisper}. 

Our dataset comprises over 1,000 selected tracks sourced from renowned international record companies. Dataset statistics are presented in Table~\ref{tab:stats}. These tracks are stored as uncompressed raw audio (\(X_{\text{raw}}\)). For each track, we only consider the drop, which is the most representative segment of the style. At times, listening to other sections like the intro could provide a rough estimate of the category, but it is the drop that determines the exact genre of a song. To identify the drop, we detect excerpts where the volume consistently remains above a threshold\footnote{$V_{\text{margin}}$ is a hyper-parameter set as 1.5dB empirically.} (\(V_{\text{thres}}= V_{\text{max}} - V_{\text{margin}}\)), where \(V_{\text{max}}\) denotes the maximum volume within the audio. Note that before detection, some rule-based smoothing process is applied to mitigate loudness fluctuations. After identifying the segments that meet the aforementioned criteria, several 7.5-second clips (around 4 bars\footnote{We extended the clip length which was 1 bar in \textit{HouseX}, since usually 4 bars make up a loop that better exhibits the style.}) are randomly sampled from each segment, ensuring a comprehensive representation of the track.

Subsequently, feature extraction is performed on each clip. Using \textit{librosa} \cite{brian_mcfee-proc-scipy-2015}, we compute: mel-spectrograms (\(X_{\text{mel}}\)), CQT-chromagrams (\(X_{\text{cqt}}\)), VQT-chromagrams (\(X_{\text{vqt}}\)), which are then stacked to produce the final audio feature matrix (\(X_{\text{feat}}\)). Detailed hyper-parameters for generating the mel-spectrogram, CQT, and VQT, are comprehensively shown in our code repository.

All tracks in the dataset are manually annotated with a total of eight distinct labels. The first four labels align with those previously defined by the \textit{HouseX} dataset \cite{9980316}. To enhance classification granularity to a level suitable for representing most contemporary mainstage live sets played in music festivals, we introduce four additional genres. Recognizing that certain tracks may exhibit characteristics of multiple genres, we employ soft labeling techniques to add more information to our labels, which gives better performance than hard 0/1 labeling, as shown in Sec. \ref{sec:results}.
\begin{table}[!ht]
\centering
\normalsize
\caption{An overview of the dataset with the size for each category along with the sample rate and the BPM (beats per minute) span.}
\begin{tabular}{lcc} 
\Xhline{2\arrayrulewidth}
\Xhline{2\arrayrulewidth}
\multirow{2}{*}{Genre} & \multicolumn{2}{c}{\# of Clips} \\
\cline{2-3}
                       & Train. & Val. \\
\hline
Progressive House & 1215 & 344 \\
Future House & 1192 & 348 \\
Bass House & 1102 & 120 \\
Tech House & 643 & 200 \\
Deep House & 591 & 116\\
Bigroom & 774 & 284 \\
Future Rave & 920 & 108 \\
Slap House & 667 & 232 \\
\hline
Total & 7104 & 1752 \\
\Xhline{2\arrayrulewidth}
\Xhline{2\arrayrulewidth}
\# of Tracks (Train. \& Val.) & \multicolumn{2}{c}{1035} \\
Sample Rate & \multicolumn{2}{c}{44100 Hz} \\
BPM Range & \multicolumn{2}{c}{115-130} \\
\Xhline{2\arrayrulewidth}
\Xhline{2\arrayrulewidth}
\end{tabular}
\label{tab:stats}
\end{table}

\begin{table*}[!htbp]
\centering
\footnotesize
\caption{Experiment results on the validation set, where the precision, the recall and the F1 score are all weighted across the 8 categories.}
\label{tab:model_performance}
\begin{threeparttable}
    \begin{tabular}{
        l ||
        S[table-format=1.3, table-align-text-post=false] S[table-format=1.3, table-align-text-post=false] |
        S[table-format=1.3, table-align-text-post=false] S[table-format=1.3, table-align-text-post=false] |
        S[table-format=1.3, table-align-text-post=false] S[table-format=1.3, table-align-text-post=false] |
        S[table-format=1.3, table-align-text-post=false] S[table-format=1.3, table-align-text-post=false] |
        S[table-format=1.3, table-align-text-post=false] S[table-format=1.3, table-align-text-post=false] |
        S[table-format=1.3, table-align-text-post=false] S[table-format=1.3, table-align-text-post=false]
    }
    \Xhline{2\arrayrulewidth} 
    \multicolumn{1}{c||}{\textbf{Architecture}} &  
    \multicolumn{4}{c|}{\textbf{Precision}\tnote{$\star$}} & 
    \multicolumn{4}{c|}{\textbf{Recall}\tnote{$\star$}} & 
    \multicolumn{4}{c}{\textbf{F1}\tnote{$\star$}} \\
    \hline
    \multicolumn{1}{c||}{chroma info\tnote{$\dagger$}} & 
    \multicolumn{2}{c}{w} & \multicolumn{2}{c|}{w/o} & 
    \multicolumn{2}{c}{w} & \multicolumn{2}{c|}{w/o} & 
    \multicolumn{2}{c}{w} & \multicolumn{2}{c}{w/o} \\
    \hline
    \multicolumn{1}{c||}{label setup\tnote{$\ddagger$}} & 
    {hard} & {soft} & {hard} & {soft} & 
    {hard} & {soft} & {hard} & {soft} & 
    {hard} & {soft} & {hard} & {soft} \\
    \hline\hline
    Ours (\textit{ViT\_B\_16}) & 
    0.077 & 0.039 & 0.039 & 0.039 & 
    0.224 & 0.199 & 0.199 & 0.199 & 
    0.109 & 0.066 & 0.066 & 0.066 \\
    Ours (\textit{VGG11\_BN}) & 
    0.746 & 0.751 & 0.739 & 0.765 & 
    0.716 & 0.748 & 0.735 & 0.757 & 
    0.722 & 0.747 & 0.731 & 0.756 \\
    Ours (\textit{DenseNet201}) & 
    0.730 & 0.767 & 0.786 & \textbf{0.790} & 
    0.712 & 0.749 & 0.752 & 0.743 & 
    0.716 & 0.741 & 0.757 & 0.751 \\
    Ours (\textit{ResNet152}) & 
    0.760 & 0.761 & 0.776 & 0.780 & 
    0.720 & 0.737 & 0.748 & \textbf{0.761} & 
    0.718 & 0.744 & 0.756 & \textbf{0.764} \\
    \Xhline{2\arrayrulewidth} 
    \textit{Qwen-Audio-Chat}$^{\S}$ & 
    \multicolumn{4}{c|}{0.122} & 
    \multicolumn{4}{c|}{0.099} & 
    \multicolumn{4}{c}{0.037} \\
    \textit{Qwen2-Audio-7B-Instruct}$^{\S}$ & 
    \multicolumn{4}{c|}{0.202} & 
    \multicolumn{4}{c|}{0.129} & 
    \multicolumn{4}{c}{0.031} \\
    \textit{Qwen2.5-Omni-7B}$^{\S}$ & 
    \multicolumn{4}{c|}{0.008} & 
    \multicolumn{4}{c|}{0.075} & 
    \multicolumn{4}{c}{0.014} \\
    \textit{Kimi-Audio-7B-Instruct}$^{\S}$ & 
    \multicolumn{4}{c|}{0.121} & 
    \multicolumn{4}{c|}{0.187} & 
    \multicolumn{4}{c}{0.109} \\
    \textit{MU-LLaMA}$^{\S}$ & 
    \multicolumn{4}{c|}{0.050} & 
    \multicolumn{4}{c|}{0.224} &
    \multicolumn{4}{c}{0.081} \\
    \textit{MusiLingo}$^{\S}$ & 
    \multicolumn{4}{c|}{0.257} & 
    \multicolumn{4}{c|}{0.283} &
    \multicolumn{4}{c}{0.116} \\
    
    \Xhline{2\arrayrulewidth} 
    \end{tabular}

    \begin{tablenotes}
        \item[$\star$] The 3 metrics here are weighted averages across all 8 genres.
        \item[$\dagger$] This row indicates whether we include chromagrams in the data representation.
        \item[$\ddagger$] This row indicates whether a hard 0/1 labeling or a soft labeling approach is used in the data representation. 
        \item[$\S$] We use GPT-4o-generated doc and human-producer expertise in the text prompt for these models, provided in the code repository.
    \end{tablenotes}
\end{threeparttable}
\end{table*}

\begin{figure*}[!htbp]
    \begin{center}
        \includegraphics[width=0.93\textwidth]{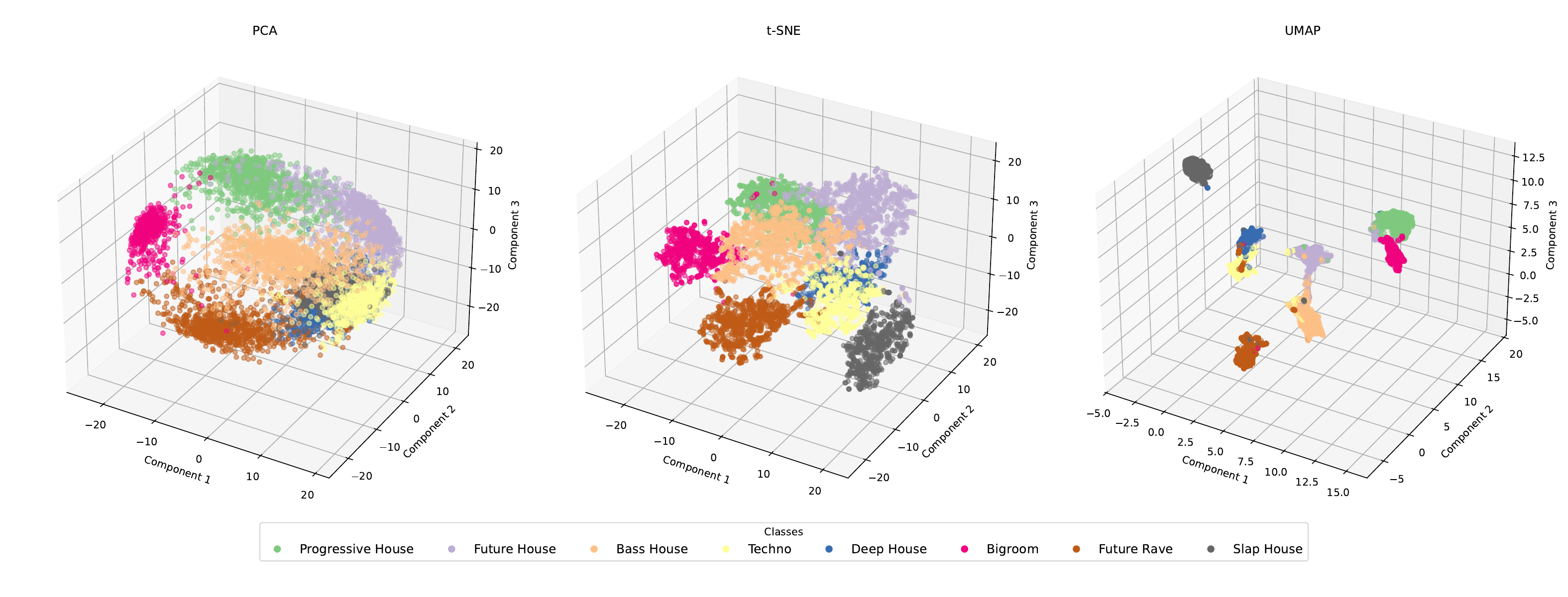} 
        \caption{3D Feature Visualization of the Training Set after Dimension Reduction}
        \label{fig:3dTrain}
    \end{center}
\end{figure*}

\subsection{Model}
\label{sub-sec:model}

Given a dataset of audio features
$\{X_{\text{feat}}^i\}_{i=1}^N$ with corresponding soft labels $p(y_i | X_{\text{feat}}^i)$ that represent probability distributions over possible genres, our objective is to train a neural network, parameterized by $\theta$, to predict genre distributions $q_\theta(y_i | X_{\text{feat}}^i)$ that closely match these soft labels.

To achieve this goal, we minimize the Kullback-Leibler (KL) divergence between the true distribution $p(y_i | X_{\text{feat}}^i)$ and the predicted distribution $q_\theta(y_i | X_{\text{feat}}^i)$:

\begin{equation}
\text{KL}(p \parallel q_\theta) = \sum_{i=1}^{N} \sum_{y_i \in \mathcal{Y}} p(y_i | X_{\text{feat}}^i) \log \left( \frac{p(y_i | X_{\text{feat}}^i)}{q_\theta(y_i | X_{\text{feat}}^i)} \right).
\end{equation}

Discarding the part that does not depend on $\theta$, this is equivalent to minimizing the cross-entropy loss:

\begin{equation}
\mathcal{L}(\theta) = - \sum_{i=1}^{N} \sum_{y_i \in \mathcal{Y}} p(y_i | X_{\text{feat}}^i) \log \left( q_\theta(y_i | X_{\text{feat}}^i) \right).
\end{equation}

Note that this formulation is consistent with cross-entropy for hard labels where $p$ simply becomes the indicator function that indicates the sole genre $X_{\text{feat}}^i$ belongs to.

The architecture of network $\theta$ is inspired by \textit{AST} (audio spectrogram transformer) \cite{gong21b_interspeech} trained on the \textit{AudioSet} \cite{audioset}. As shown in column (b) in Figure~\ref{fig:pipeline}, we derive a sequence of overlapping patches\footnote{$M$ is determined by the hyper-parameter ``hop length".} $\{X_{\text{feat}}^{i,j}\}_{j=1}^M$ from each acoustic feature matrix $X_{\text{feat}}^i$, where $X_{\text{feat}}^{i,j}$ is square with shape ($3\times224\times224$). Then each square feature patch is fed into a vision encoder, which essentially consists of conventional architectures used for image classification \cite{He2015DeepRL,Huang2016DenselyCC,Simonyan2014VeryDC,Dosovitskiy2020AnII}. The output for the sequence of input feature patches is a corresponding sequence of embeddings $\{Z_{\text{emb}}^{i,j}\}_{j=1}^M$. Positional encodings are then added to this sequence of embeddings, which are subsequently passed through several stacked transformer \cite{attention_is_all_you_need} encoder layers. The first (on the $seq\_len$ axis) output from this transformer is finally fed into a linear layer to obtain the predicted distribution.

Compared with the previous architecture in \textit{HouseX} \cite{9980316}, which is solely an image classifier, we convert the raw feature matrix into a sequence of patches to accommodate the quadrupled clip length. In fact, keep resizing the feature matrix into squares in a brute-force manner would lead to severe information loss on the temporal axis, which is crucial for characterizing genres like future house that contain very fragmentary music notes or samples.

\section{Results}
\label{sec:results}

This section presents the experiment results alongside our interpretations and findings. Table~\ref{tab:model_performance} provides numerical details and Figure~\ref{fig:3dTrain} shows the final embedding space after feature reduction by PCA, t-SNE \cite{JMLR:v9:vandermaaten08a} and UMAP \cite{mcinnes2020umapuniformmanifoldapproximation}. By convention, precision, recall, and F1-score were computed after sharpening the soft labels into hard categories (e.g., converting a soft label of [$0.25, 0.75, 0, \ldots$] into [$0, 1, 0, \ldots$]) for evaluation purposes. This approach allows the results to be interpreted as the model's ability to correctly identify the primary characteristic of a song.

Four popular CNN/ViT architectures (with comparable number of parameters) serve as the feature extractor. Experiments show that all our CNN-based specialized models outperform the rest significantly. For now, the lower performance of ViT-based methods is attributed to the need of high temporal resolution for EDM genre classification, where CNN architectures inherently provide stronger inductive biases suited for audio spectrogram data. Again, our purpose is to justify the necessity of building specific dataset rather than ``beat" LLM-based models. In other words, MLLMs are quite likely to perform well when provided with proper post-training data; otherwise, they struggle to align textual background information with domain-specific audio features. In addition, models trained on soft labels perform uniformly better than those trained from (sharpened) 0/1 labels, which supports our claim that soft labeling provides richer information of the tracks. Moreover, models trained on composite data with chromagrams fail to surpass those trained solely on mel-spectrogram. We attribute this phenomenon to the domain gap between the RGB space of \textit{ImageNet} \cite{imagenet} and the mel-CQT-VQT space. Future work may explore refining network architectures or leveraging existing music encoders--such as \textit{MERT} \cite{li2023mert}, \textit{MusicFM} \cite{won2023foundationmodelmusicinformatics}, or \textit{MuQ} \cite{zhu2025muqselfsupervisedmusicrepresentation}--which are trained on broader musical genres, to improve performance on our composite dataset.

We also extract the embeddings before the final linear layer and provide visualization with dimension reduction techniques. The figures indicate that progressive house, bigroom and slap house are relatively well-separated, aligning with our annotations where most non-0/1 labels involve other genres.

For hardware configuration, at least 24G GPU memory is required to complete one run with a batch size of 4. We assign one NVIDIA RTX A6000 GPU to each data setting (whether or not to use chromagram and whether or not to use soft labels).

\section{Applications}
\label{sec:application}

We propose some real-world scenarios for such classification algorithm. A straightforward application is music recommendation system tailored for listeners with specific preferences on certain sub-genres. Besides, this algorithm can boost productivity in multimedia contexts, such as automated MV generation and visuals generation given certain pre-defined rules \cite{9980316}. For illustrative purposes, we prototyped a visual automation system simulated in Blender 3D\footnote{Demos are available in the links given in the code repository.}.

\section{Conclusion}
\label{sec:conclusion}

We developed a comprehensive house music classification benchmark with 8 classes to address the limitations of existing datasets, such as limited sub-genre representation and inter-class overlaps. Our dataset employs continuous soft labeling to better capture track characteristics. Our study highlights the inadequacy of existing machine learning methods in understanding sub-genres of electronic music, thus calling for the development of specialized datasets and models.

Future work may explore several promising directions. One avenue is to further scale up the dataset to better utilize the CQT and VQT feature spaces. Since labeling large datasets is impractical for the machine learning community alone, a collaborative approach with producers/songwriters from the music industry is recommended. Additionally, future MIR research should also extend beyond sub-genre classification to encompass timbral and rhythmic characteristics, which retrieves more information that could benefit downstream scenes. On the other hand, training an MLLM capable of captioning EDM tracks with descriptive attributes remains a challenging but promising task to empower downstream applications.


\bibliographystyle{IEEEtran}
\bibliography{refs25}








\end{document}